# Magnetization dynamics of two interacting spins in an external magnetic field

Sergey V. Titov,[a] Hamid Kachkachi,[b] Yuri P. Kalmykov,[c] and William T. Coffey [d]

[a] *Institute of Radio Engineering and Electronics of the Russian Academy of Sciences, Fryazino, Moscow Region, 141190, Russian Federation*

[b] *Laboratoire de Magnétisme et d'Optique, Université de Versailles St. Quentin 45 av. des Etats-Unis, 78035 Versailles Cedex, France*

[c] *Laboratoire de Mathématiques et Physique des Systèmes, Université de Perpignan, 52 Avenue Paul Alduy, 66860 Perpignan Cedex, France*

[d] *Department of Electronic and Electrical Engineering, Trinity College, Dublin 2, Ireland*

**Abstract**

The longitudinal relaxation time of the magnetization of a system of two exchange coupled spins subjected to a strong magnetic field is calculated exactly by averaging the stochastic Gilbert-Landau-Lifshitz equation for the magnetization, i.e., the Langevin equation of the process, over its realizations so reducing the problem to a system of linear differential-recurrence relations for the statistical moments (averaged spherical harmonics). The system is solved in the frequency domain by matrix continued fractions yielding the complete solution of the two spin problem in external fields for all values of the damping and barrier height parameters. The magnetization relaxation time extracted from the exact solution is compared with the inverse relaxation rate from Langer's theory of the decay of metastable states [J. S. Langer, Ann. Phys. (N.Y.) **54**, 258 (1969)], which yields in the high barrier and intermediate-to-high damping limits the asymptotic behavior of the greatest relaxation time.

PACS number(s): 75.50.Tt, 76.20.+q, 05.40.Jc

# I. INTRODUCTION

Fine single-domain ferromagnetic particles are characterized by thermal instability of their magnetization due to thermal agitation.[1] Thermal fluctuations and relaxation of the magnetization of single domain particles are important in information storage and rock magnetism, as well as in magnetization reversal (that is the slowest magnetization decay mode which is characterized by the greatest relaxation time) in isolated ferromagnetic nanoparticles and nanowires. The effect of thermal fluctuations on the magnetization reversal of an assembly of single domain ferromagnetic particles (that is single coherent spins) is described by the Néel-Brown model.[1-3] This model, based on the classical theory of the Brownian motion (by taking as Langevin equation the Gilbert-Landau-Lifshitz equation for the motion of the magnetization augmented by a random field) and the one-spin approximation, has been used to interpret many experiments (see, for example, Ref. 4). Nevertheless, systems of (both metallic and ferrite) particles exist, where deviations in behavior from that predicted by the one-spin approximation have been recently observed.[5-8] An understanding of these effects requires microscopic theories of the magnetization dynamics, capable of distinguishing and accounting for the various crystallographic local environments inside a nanoparticle and on its surface. In formulating such theories one is invariably faced with complex *N*-body problems. Thus before one can calculate the magnetization relaxation time of many interacting spins, one needs to understand the effect of interactions on the magnetization relaxation time of the simplest interacting system, conceived of as a pair of spins coupled via exchange interaction, including the usual magneto-crystalline anisotropy and Zeeman terms. This will be referred to as the *two-spin problem* (TSP). This apparently simple problem presents, in the course of its general formulation, the usual difficulties related with analyzing the energyscape (location of the minima, maxima, and saddle points of the potential energy) in a system with many degrees of freedom, which is a crucial step in the calculation of the relaxation time whatever the context. We remark that the solution of the TSP is also of interest in the study of the dielectric relaxation of polar molecules containing rotating polar groups interacting by means of dipole-dipole coupling as formulated by Budó[9] and which has been reviewed in detail in Ref. 10.

The effect of interactions between magnetic moments on the magnetization relaxation time $\tau$ has been treated in several papers. For example, Lyberatos *et al.*[11] used a Monte-Carlo method to study the dynamics of an assembly of interacting magnetic moments. In Ref. 12, a pair of coupled dipoles was treated using numerical Langevin dynamics simulations. Hinzke and Nowak[13,14] compared calculations of thermally activated magnetization reversal in systems of interacting classical magnetic moments both by a Monte-Carlo method and Langevin dynamics simulations. Denisov and Trohidou[15] by solving the Fokker-Planck equation for the distribution function of two-dimensional ensembles of ferromagnetic nanoparticles, derived an evolution equation for magnetization and found its solution in certain limiting cases. Approximate analytic equations for $\tau$ were obtained under particular conditions for assemblies of magnetic moments in Refs. 16-18. Rodé *et al.*[19] numerically solved the Fokker-Planck equation for a system of two interacting particles and calculated the time decay of the magnetization. Chen *et al.*[20] presented an analytic solution for $\tau$ for two identical interacting single domain particles and obtained good agreement with the numerical results using the Fokker-Planck equation. Yoshimori and Korringa[22] and Solomon[21] treated relaxational processes of a system of two spins in a magnetic field. In the context of superparamagnetism, the relaxation time for two equivalent spins in a uniaxial potential and a magnetic field has been evaluated by Kachkachi[23] using Langer's general theory of the decay of metastable states.[24] This theory also comprises the generalization of Kramers calculation of the intermediate-to-high damping (IHD) escape rate for a single degree of freedom system with a separable and additive Hamiltonian to multidegree of freedom systems with non separable Hamiltonians as obtain in magnetic systems. The IHD solution corresponds physically to the situation where the energy loss per cycle of the almost periodic motion of a spin having the saddle point energy is $\geq kT$. In this limit, Kachkachi[23] considered the TSP in the special situation, where the easy axes of the two spins are parallel to each other and to the applied field. He then reformulated Langer's general reaction rate equation and used it to obtain analytical expressions for the relaxation time (a detailed description of the application of Langer's theory of the decay of metastable states to superparamagnetism and its relation to the Kramers escape rate is given in Refs. 25-27).

The purpose of the present work is to present the exact solution of the TSP (a system with more than two degrees of freedom). This will be accomplished by (i) deriving an exact system of equations for statistical moments by directly averaging the Gilbert-Landau-Lifshitz equations for the motion of the magnetization augmented by a random field due to the heat bath over its realizations; (ii) solving the resulting hierarchy of moment equations by the matrix-continued fraction (MCF) method.[26,28] This method will allow us to evaluate the time decay of the magnetization in all other ranges of the damping, exchange coupling, anisotropy and applied field parameters as well as other physical parameters (spectra of the relaxation functions, the complex susceptibility, etc.). In the IHD regime, asymptotic formulae for the greatest relaxation time[23] based on Langer's calculation are naturally expected to apply.[27] In this regime the results for the relaxation time obtained using the exact MCF method are compared with those of Ref. 23 in order to establish a range of validity for asymptotic calculations of the IHD relaxation time.

## II. SOLUTION OF THE LANGEVIN EQUATIONS FOR TWO INTERACTING SPINS

We first demonstrate how the hierarchy of differential-recurrence relations for the appropriate relaxation functions arise naturally from the coupled vector Langevin equations for the two-spin system thus bypassing the problem of constructing and solving the Fokker-Planck equation entirely. Consider a system of two exchange-coupled spins $\mathbf{S}_p(t)$, $p=1,2$,

$$\mathbf{S}_p = S\mathbf{s}_p, \quad \mathbf{s}_p = \mathbf{i}\sin\vartheta_p\cos\varphi_p + \mathbf{j}\sin\vartheta_p\sin\varphi_p + \mathbf{k}\cos\vartheta_p, \tag{1}$$

where $\vartheta_p$ and $\varphi_p$ are the respective polar and azimuthal angles, $\mathbf{s}_p$ the unit vectors along $\mathbf{S}_p$, and $S$ the nominal value of the spin. The two easy axes are taken parallel to each other and to the applied field, which is parallel to the reference ($Z$) axis. Next, we assume that the magnitude of an external applied (spatially) uniform dc magnetic field is suddenly altered at time $t = 0$ from $\mathbf{H}_Z^{I}$ to $\mathbf{H}_Z^{II}$. We are then interested in the relaxation of this system starting from an equilibrium state I with the Boltzmann distribution function $W_I = Z_I^{-1} e^{-\beta V_I}$ ($t \leq 0$) to another equilibrium state II with the distribution function $W_{II} = Z_{II}^{-1} e^{-\beta V_{II}}$ ($t \to \infty$). Here $V_\gamma$ is the free energy, $\beta = 1/kT$ and $Z_\gamma$ ($\gamma =$ I, II) are the partition functions. The dynamics of the spin $\mathbf{S}_p(t)$ immediately following the alteration of the field may be described using the normalized relaxation function

$$f_p(t) = \frac{\langle \mathbf{s}_p \cdot \mathbf{e}_Z \rangle(t) - \langle \mathbf{s}_p \cdot \mathbf{e}_Z \rangle_{II}}{\langle \mathbf{s}_p \cdot \mathbf{e}_Z \rangle_I - \langle \mathbf{s}_p \cdot \mathbf{e}_Z \rangle_{II}}, \tag{2}$$

where $\mathbf{e}_Z$ is the unit vector along the Z axis, the angular brackets $\langle \mathbf{s}_p \cdot \mathbf{e}_Z \rangle(t)$ are the time dependent ensemble averages, and the brackets $\langle \otimes \rangle_\gamma$ designate the *equilibrium* ensemble average in the initial I and final II states,

$$\langle \otimes \rangle_\gamma = \int_0^{2\pi} \int_0^{2\pi} \int_0^{\pi} \int_0^{\pi} \otimes W_\gamma(\vartheta_1, \varphi_1, \vartheta_2, \varphi_2) \sin \vartheta_2 \sin \vartheta_1 d\vartheta_2 d\vartheta_1 d\varphi_2 d\varphi_1. \tag{3}$$

Note that the transient response so formulated is truly *nonlinear* because the change in amplitude $H_I - H_{II}$ of the external dc magnetic field is *arbitrary*. The overall behavior of $f_p(t)$ is characterized by the *integral relaxation time* $\tau$ [26]. This time is the area under the curve of $f_p(t)$ [26]

$$\tau = \int_0^\infty f_1(t) dt = \int_0^\infty f_2(t) dt. \tag{4}$$

Here for values of the external applied field less than a certain critical field (see below) $\tau$ yields a close approximation to the greatest relaxation time of the system. The relaxation function $f_1(t)$ [or $f_2(t)$] and the relaxation time $\tau$ also describes the transient nonlinear behavior of the longitudinal component of the magnetic dipole moment of the system $m_Z(t) = \mu \langle (\mathbf{s}_1 + \mathbf{s}_2) \cdot \mathbf{e}_Z \rangle(t)$ [which is proportional to the magnetization $M_Z(t)$], viz.,

$$m_Z(t) = 2\mu \left[ \langle \mathbf{s}_1 \cdot \mathbf{e}_Z \rangle_{II} + \left( \langle \mathbf{s}_1 \cdot \mathbf{e}_Z \rangle_I - \langle \mathbf{s}_1 \cdot \mathbf{e}_Z \rangle_{II} \right) f_1(t) \right], \tag{5}$$

where $\mu = g\mu_B S$ is the magnetic moment associated with the spin $\mu_B$ is the Bohr magneton, and $g$ is the Landé factor. Moreover, one can also evaluate the *linear response* of a two-spin system to infinitesimally small changes in the strength of the strong dc field $\mathbf{H}_Z^I$, i.e., for $\mathbf{H}_Z^{II} = \mathbf{H}_Z^I - \mathbf{\kappa}$ as $\mathbf{\kappa} \to \mathbf{0}$, where $\mathbf{\kappa}$ is considered as a small external perturbation. Here the relaxation function $f_p(t)$ from Eq. (2) coincides with the normalized longitudinal dipole equilibrium correlation function $C(t)$, that is

$$\lim_{\kappa \to 0} f_p(t) = C(t) = \frac{\langle m_Z(0) m_Z(t) \rangle_{II} - \langle m_Z(0) \rangle_{II}^2}{\langle m_Z^2(0) \rangle_{II} - \langle m_Z(0) \rangle_{II}^2}. \tag{6}$$

According to linear response theory (see, e.g., [26]), having determined the one-sided Fourier transform $\tilde{C}(\omega) = \int_0^\infty C(t) e^{-i\omega t} dt$ [the spectrum of the equilibrium correlation function $C(t)$], one can calculate the integral relaxation time in the linear response approximation that is the correlation time $\tau = \tilde{C}(0)$ and the normalized dynamic susceptibility $\hat{\chi}(\omega) = \chi'(\omega) - i\chi''(\omega)$ [26]

$$\hat{\chi}(\omega) = 1 - i\omega \tilde{C}(\omega). \tag{7}$$

The asymptotic behavior of $\hat{\chi}(\omega)$ in the extreme cases of very low and very high frequencies is given by[26]

$$\hat{\chi}(\omega) \begin{cases} 1 - i\omega \int_0^\infty C(t) dt = 1 - i\omega\tau, & \omega \to 0, \\ \dfrac{\dot{C}(0)}{i\omega} + ... = -\dfrac{i}{\omega \tau^{ef}} + ..., & \omega \to \infty, \end{cases} \tag{8}$$

where

$$\tau^{ef} = -1/\dot{C}(0) \tag{9}$$

is the effective relaxation times which will be given later on. The times $\tau$ and $\tau^{ef}$ also characterizes the long and short time behavior of $C(t)$, respectively.[26]

The behavior of $f_p(t)$, $\tau$, $C(t)$, and $\hat{\chi}(\omega)$ is completely determined by the spin dynamics, which is governed by the stochastic Landau-Lifshitz-Gilbert equation, i.e., the deterministic Landau-Lifshitz-Gilbert equation augmented by a random field $\mathbf{h}^{(p)}(t)$,[3,26]

$$\dot{\mathbf{s}}_p(t) = b\mu\left(\alpha^{-1}\left[\mathbf{s}_p(t)\times\left(\mathbf{H}_p(t)+\mathbf{h}_p(t)\right)\right] - \left[\mathbf{s}_p(t)\times\left[\mathbf{s}_p(t)\times\left(\mathbf{H}_p(t)+\mathbf{h}_p(t)\right)\right]\right]\right), \quad (10)$$

where $b=\beta/(2\tau_N)$, $\tau_N = \beta\mu(1+\alpha^2)/(2\gamma\alpha)$ is a characteristic (free diffusion) time, $\gamma$ is the gyromagnetic ratio, $\alpha$ is the dimensionless damping parameter representing the dissipative coupling to the heat bath and the random Gaussian field $\mathbf{h}_p(t)$ has the white noise properties

$$\overline{h_{\vartheta_p}(t)} = \overline{h_{\varphi_p}(t)} = 0, \quad \overline{h_{\vartheta_p}(t)h_{\vartheta_p}(t')} = \overline{h_{\varphi_p}(t)h_{\varphi_p}(t')} = 2\alpha(\gamma\mu\beta)^{-1}\delta(t-t'). \quad (11)$$

Here the overbar stands for statistical averaging over an ensemble of spins which have all started with the same (sharp) values of $\mathbf{s}_p(t) = \mathbf{s}_p$. The random field takes into account the thermal fluctuations of the spin via the fluctuation-dissipation theorem. The effective magnetic field $\mathbf{H}_p(t)$ acting on the spin $p$ consists of the externally applied magnetic fields, the crystalline anisotropy field, and the molecular (or exchange) field produced by the other spin. It may be written as

$$\mathbf{H}_p = \frac{1}{\mu}\left(0, \ -\frac{\partial V}{\partial \vartheta_p}, \ -\frac{1}{\sin\vartheta_p}\frac{\partial V}{\partial \varphi_p}\right). \quad (12)$$

The reduced free energy $\beta V$ of the two spin system in the dc magnetic field may be written as[23]

$$\beta V = -\varsigma(\mathbf{s}_1 \cdot \mathbf{s}_2) - \sum_{p=1,2}\left[\xi_\gamma\left(\mathbf{e}_Z \cdot \mathbf{s}_p\right) + \sigma\left(\mathbf{e}_Z \cdot \mathbf{s}_p\right)^2\right]. \quad (13)$$

where $\xi_\gamma = \beta\mu H_Z^\gamma$, $\sigma = \beta K$, and $\varsigma = \beta J S^2$ are the dimensionless field, anisotropy, and interaction parameters, respectively, $K$ and $J$ are the anisotropy and ferromagnetic exchange coupling constants. The vector stochastic differential equation (10) written in spherical polar coordinates leads to the system of two coupled scalar stochastic differential equations

$$\dot{\vartheta}_p(t) = b\mu\left[h_{\vartheta_p}(t) - \alpha^{-1}h_{\varphi_p}(t)\right] - b\left[\frac{\partial V(t)}{\partial \vartheta_p} - \frac{1}{\alpha \sin\vartheta_p(t)}\frac{\partial V(t)}{\partial \varphi_p}\right], \quad (14)$$

$$\dot{\varphi}_p(t) = \frac{b\mu}{\sin\vartheta_p(t)}\left[h_{\varphi_p}(t) + \alpha^{-1}h_{\vartheta_p}(t)\right] - b\left[\frac{1}{\sin^2\vartheta_p(t)}\frac{\partial V(t)}{\partial \varphi_p} + \frac{1}{\alpha \sin\vartheta_p(t)}\frac{\partial V(t)}{\partial \vartheta_p}\right]. \quad (15)$$

In applications to spin reorientational dynamics, the relevant quantities are averages involving the spherical harmonics $Y_{l,m}(\vartheta,\varphi)$, which are defined as[29]

$$Y_{l,m}(\vartheta,\varphi) = (-1)^m \sqrt{\frac{(2l+1)(l-m)!}{4\pi(l+m)!}} e^{im\varphi} P_l^m(\cos\vartheta), \quad |m| \le l$$

where $P_l^m(x)$ are the associated Legendre functions. We now introduce the functions

$$M_{l_1,l_2,m}(t) = Y_{l_1,m}[\vartheta_1(t),\varphi_1(t)]Y_{l_2,-m}[\vartheta_2(t),\varphi_2(t)], \quad (16)$$

corresponding a complete set of orthogonal functions characterizing the dynamics of the two exchange-coupled spins. Thus we can obtain from Eqs. (14)-(16) the stochastic equation of motion for the functions $M_{l_1,l_2,m}(t)$

$$\frac{d}{dt}M_{l_1,l_2,m}(t) = \sum_{p=1,2}\dot{\vartheta}_p(t)\frac{\partial M_{l_1,l_2,m}}{\partial \vartheta_p} + \dot{\varphi}_p(t)\frac{\partial M_{l_1,l_2,m}}{\partial \varphi_p}. \quad (17)$$

Upon averaging the stochastic differential Eq. (17) over its realizations (see details in Ref. 26, Chapter 7), we obtain

$$2\tau_N \dot{M}_{l_1,l_2,m} = \sum_{p=1,2}\frac{\beta}{2}\left[\left(L^{(p)}\right)^2\left(VM_{l_1,l_2,m}\right) - V\left(L^{(p)}\right)^2 M_{l_1,l_2,m} - M_{l_1,l_2,m}\left(L^{(p)}\right)^2 V\right]$$
$$-\sum_{p=1,2}\frac{i\beta}{2\alpha}\sqrt{\frac{3}{2\pi}}\left\{\left(Y_{1,1}^{(p)}\right)^{-1}\left[\left(L_Z^{(p)}V_+^{(p)}\right)\left(L_+^{(p)}M_{l_1,l_2,m}\right) - \left(L_+^{(p)}V_+^{(p)}\right)\left(L_Z^{(p)}M_{l_1,l_2,m}\right)\right]\right. \quad (18)$$
$$\left. + \left(Y_{1,-1}^{(p)}\right)^{-1}\left[\left(L_Z^{(p)}V_-^{(p)}\right)\left(L_-^{(p)}M_{l_1,l_2,m}\right) - \left(L_-^{(p)}V_-^{(p)}\right)\left(L_Z^{(p)}M_{l_1,l_2,m}\right)\right]\right\} - \sum_{p=1,2}\left(L^{(p)}\right)^2 M_{l_1,l_2,m},$$

where

$$\left(L^{(p)}\right)^2 = -\frac{1}{\sin\vartheta_p}\frac{\partial}{\partial\vartheta_p}\left(\sin\vartheta_p\frac{\partial}{\partial\vartheta_p}\right) - \frac{1}{\sin^2\vartheta_p}\frac{\partial^2}{\partial\varphi_p^2}, \quad L_Z^{(p)} = -i\frac{\partial}{\partial\varphi_p}, \quad L_\pm^{(p)} = e^{\pm i\varphi_p}\left(\pm\frac{\partial}{\partial\vartheta_p} + i\cot\vartheta_p\frac{\partial}{\partial\varphi_p}\right)$$

are the orbital angular momentum operators[29]. Here we have used the following representation for the expansion of $V$, Eq. (13), in terms of spherical harmonics

$$\beta V = -\frac{4\pi}{3}\varsigma\sum_{m=-1}^{1}(-1)^m Y_{1,m}(\vartheta_1,\varphi_1)Y_{1,m}(\vartheta_2,\varphi_2) - \sum_{p=1,2}\left[\xi_N\sqrt{\frac{4\pi}{3}}Y_{1,0}(\vartheta_p,\varphi_p) + \sigma\frac{4}{3}\sqrt{\frac{\pi}{5}}Y_{2,0}(\vartheta_p,\varphi_p)\right] + \text{Const}, \quad (19)$$

$$V = V_+^{(p)} + V_-^{(p)}, \qquad V_+^{(p)} = \sum_{R=1}^{2}\sum_{S=0}^{1} v_{R,S}^{(p)} Y_{R,S}^{(p)}, \qquad V_-^{(p)} = v_{1,-1}^{(p)} Y_{1,-1}^{(p)},$$

Noting the properties of the angular momentum operators[29]

$$\left(L^{(p)}\right)^2 Y_{l,m}(\vartheta_p,\varphi_p) = l(l+1)Y_{l,m}(\vartheta_p,\varphi_p), \quad L_Z^{(p)} Y_{l,m}(\vartheta_p,\varphi_p) = m Y_{l,m}(\vartheta_p,\varphi_p),$$

$$L_\pm^{(p)} Y_{l,m}(\vartheta_p,\varphi_p) = \sqrt{l(l+1) - m(m\pm 1)}\, Y_{l,m\pm 1}(\vartheta_p,\varphi_p),$$

Eq. (18) can be further transformed to the moment system:

$$\tau_N \dot{M}_{l_1,l_2,m} = \sum_{i,j=-2}^{2}\sum_{k=-1}^{1} d_{l_1+i,l_2+j,m+k}^{l_1,l_2,m} M_{l_1+i,l_2+j,m+k}, \qquad (20)$$

where the coefficients $d_{l_1+i,l_2+j,m+k}^{l_1,l_2,m}$ are given in Appendix A. Taking the equilibrium ensemble averages over the sharp values of $\vartheta_p$ and $\varphi_p$ [26], we now have from Eq. (20) the set of recurrence relations for the relaxation functions $c_{l_1,l_2,m}(t) = \langle M_{l_1,l_2,m}\rangle(t) - \langle M_{l_1,l_2,m}\rangle_{\text{II}}$, viz.,

$$\tau_N \dot{c}_{l_1,l_2,m} = \sum_{i,j=-2}^{2}\sum_{k=-1}^{1} d_{l_1+i,l_2+j,m+k}^{l_1,l_2,m} c_{l_1+i,l_2+j,m+k}. \qquad (21)$$

This system of differential recurrence relations for the observables must be solved subject to the initial conditions $c_{l_1,l_2,m}(0) = \langle M_{l_1,l_2,m}\rangle_{\text{I}} - \langle M_{l_1,l_2,m}\rangle_{\text{II}}$. In order to derive Eq. (21), one should note that the equilibrium averages $\langle M_{l_1,l_2,m}\rangle_\gamma$ ($\gamma$ = I, II) satisfy the recurrence relation:

$$\sum_{i,j=-2}^{2}\sum_{k=-1}^{1} d_{l_1+i,l_2+j,m+k}^{l_1,l_2,m} \langle M_{l_1+i,l_2+j,m+k}\rangle_\gamma = 0. \qquad (22)$$

The equilibrium averages $\langle M_{l_1,l_2,m} \rangle_\gamma$ can also be evaluated directly using Eq. (3).

## III. MATRIX CONTINUED FRACTION SOLUTION

To proceed we first introduce the vectors

$$\mathbf{C}_n(t) = \begin{pmatrix} \mathbf{c}_{2n-1,0}(t) \\ \mathbf{c}_{2n-2,1}(t) \\ \vdots \\ \mathbf{c}_{0,2n-1}(t) \\ \mathbf{c}_{2n,0}(t) \\ \mathbf{c}_{2n-1,1}(t) \\ \vdots \\ \mathbf{c}_{0,2n}(t) \end{pmatrix}_{4n^2+2n+1}, \quad \mathbf{c}_{n,m}(t) = \begin{pmatrix} c_{n,m,-r}(t) \\ c_{n,m,-r+1}(t) \\ \vdots \\ c_{n,m,r}(t) \end{pmatrix}, \quad r = \min[n,m], \tag{23}$$

Thus Eq. (20) can be transformed into a tridiagonal vector recurrence relation of the form

$$\tau_N \frac{d}{dt}\mathbf{C}_n(t) = \mathbf{Q}_n^- \mathbf{C}_{n-1}(t) + \mathbf{Q}_n \mathbf{C}_n(t) + \mathbf{Q}_n^+ \mathbf{C}_{n+1}(t), \tag{24}$$

with $\mathbf{C}_0(t) = \mathbf{0}$. The matrices $\mathbf{Q}_n, \mathbf{Q}_n^+, \mathbf{Q}_n^-$ are given in Appendix A. By using the general method for solving matrix recursion Eq. (24),[26] we have the exact solution for the spectrum $\tilde{\mathbf{C}}_1(\omega)$ as

$$\tilde{\mathbf{C}}_1(\omega) = \tau_N \Delta_1(\omega) \left\{ \mathbf{C}_1(0) + \sum_{n=2}^\infty \left( \prod_{k=2}^n \mathbf{Q}_{k-1}^+ \Delta_k(\omega) \right) \mathbf{C}_n(0) \right\}, \tag{25}$$

where the matrix continued fraction $\Delta_n(\omega)$ is defined by the recurrence equation

$$\Delta_n(\omega) = \left[ i\omega\tau_N \mathbf{I} - \mathbf{Q}_n - \mathbf{Q}_n^+ \Delta_{n+1}(\omega) \mathbf{Q}_{n+1}^- \right]^{-1}$$

and the tilde denotes the one-sided Fourier transform. The initial vector $\mathbf{C}_n(0)$ in Eq. (25) can also be calculated in terms of matrix continued fractions (see Appendix A).

Having determined $\tilde{\mathbf{C}}_1(\omega)$, one can evaluate the spectrum of the relaxation function $f_p(t)$

$$\tilde{f}_p(\omega) = \tilde{c}_{1,0,0}(\omega)/c_{1,0,0}(0) \tag{26}$$

and thus the spectrum of the magnetization of the system $M_Z(t)$ [see Eq. (5)] and its relaxation time $\tau$ given by the zero frequency limit of $\tilde{f}_p(\omega)$ [cf. Eq. (4)]

$$\tau = \tilde{c}_{1,0,0}(0)/c_{1,0,0}(0). \tag{27}$$

For linear response, one can also calculate from Eqs. (6) and (7) the spectrum of the equilibrium correlation function $C(t)$, the correlation time $\tau = \tilde{C}(0)$, and the longitudinal complex susceptibility $\chi(\omega) = \chi'(\omega) - i\chi''(\omega)$. The time behavior of $f_p(t)$, $C(t)$, and $M_Z(t)$ can be obtained numerically from $\tilde{f}_p(\omega)$ and $\tilde{C}(\omega)$ by inverse Fourier transformation.

The integral relaxation (correlation) time $\tau$ and effective relaxation time $\tau^{ef}$ may equivalently be written in terms of the eigenvalues ($\lambda_k$) of the Fokker-Planck operator $L_{FP}$ as defined by the Fokker-Planck equation underlying Eq. (10), viz.

$$\dot{W} = L_{FP} W, \tag{28}$$

for the distribution function $W(\mathbf{s}_1, \mathbf{s}_2, t)$ of the orientations of the spins. The latter definition follows because the function $f_p(t)$ may formally be written as the discrete set of relaxation modes[26,28]

$$C(t) = \sum_k c_k e^{-\lambda_k t}, \tag{29}$$

where $\sum_k c_k = 1$, so that from Eqs. (9), (27)-(29)

$$\tau = \sum_k c_k/\lambda_k \quad \text{and} \quad \tau^{ef} = \left(\sum_k c_k \lambda_k\right)^{-1}. \tag{30}$$

Equation (30) emphasizes that the relaxation times $\tau$ and $\tau^{ef}$ contain contributions from *all* the eigenvalues $\lambda_k$. The smallest nonvanishing eigenvalue $\lambda_1$ is associated with the slowest overbarrier or interwell relaxation mode and so with the long-time behavior of $f_p(t)$; the other eigenvalues $\lambda_k$ characterize high-frequency "intrawell" modes. In general, in order to evaluate $\tau$ numerically from Eq. (30), all $\lambda_k$ and $c_k$ are required. However, in the low temperature (high barrier) limit, $\lambda_1 << |\lambda_k|$ and $c_1 \approx 1 >> c_k$ ($k \neq 1$) provided the wells of the potential remain approximately equivalent (as is true for a small external field) so that

$$\tau \approx 1/\lambda_1. \tag{31}$$

Thus the integral relaxation time $\tau$ for nearly equivalent potential wells in the low temperature limit closely approximates $\lambda_1^{-1}$, that is the greatest relaxation time of the magnetization. By using matrix continued fractions, one can also estimate the smallest nonvanishing eigenvalue $\lambda_1$ of the Fokker-Planck operator $L_{FP}$, from the secular equation[26]

$$\det\left[\lambda_1 \tau_N \mathbf{I} + \mathbf{Q}_1 + \mathbf{Q}_1^+ \mathbf{\Delta}_2(-\lambda_1) \mathbf{Q}_2^-\right] = 0. \tag{32}$$

We remark that for linear response $\lambda_1$ can also be evaluated from the half-width of the spectrum $\tilde{C}(\omega)$ or, equivalently, from the low-frequency maximum of the loss spectra $\chi''(\omega)$.

The effective relaxation time $\tau^{ef}$, Eq. (9), may also be evaluated from the averaged Gilbert equations (14) and (15) in terms of equilibrium averages only. We have

$$\tau_N \frac{d}{dt}\cos\vartheta_p = -\cos\vartheta_p + \frac{\beta \sin\vartheta_p}{2}\frac{\partial V}{\partial \vartheta_p} - \frac{\beta}{2\alpha}\frac{\partial V}{\partial \varphi_p}, \quad (p=1,2). \tag{33}$$

The first term in the right hand side of Eq. (33) is the noise-induced drift. Equations (33) can also be obtained from Eq. (20). It follows from Eq. (33) that

$$\left\langle (\cos\vartheta_1 + \cos\vartheta_2)\frac{d}{dt}(\cos\vartheta_1 + \cos\vartheta_2) \right\rangle_{II}$$
$$= -\frac{1}{\tau_N}\left\langle (\cos\vartheta_1 + \cos\vartheta_2)\sum_{p=1}^{2}\left(\cos\vartheta_p - \frac{\beta \sin\vartheta_p}{2}\frac{\partial V}{\partial \vartheta_p} + \frac{\beta}{2\alpha}\cos\vartheta_p\frac{\partial V}{\partial \varphi_p}\right) \right\rangle_{II} \tag{34}$$
$$= -\frac{1}{2\tau_N}\left\langle (\sin^2\vartheta_1 + \sin^2\vartheta_2) \right\rangle_{II}$$

(here we have used integration by parts). Thus, according to Eq. (9) we have

$$\tau^{ef} = 2\tau_N \frac{\left\langle (\cos\vartheta_1 + \cos\vartheta_2)^2 \right\rangle_{II} - \left\langle \cos\vartheta_1 + \cos\vartheta_2 \right\rangle_{II}^2}{\left\langle \sin^2\vartheta_1 + \sin^2\vartheta_2 \right\rangle_{II}}. \quad (35)$$

The equilibrium averages (quadruple integrals) defined by Eq. (3) can be evaluated numerically.

We remark that the two-spin problem has been treated in the particular situation of the two easy axes parallel to each other and to the applied field. The general situation of an arbitrary angle between the easy axes and different anisotropy constants can be analyzed in like manner. Here the reduced free energy $\beta V$, Eq. (19), contains spherical harmonics of the first and second order only so that averaging of the relevant stochastic equations yields a recurrence relation for the statistical moments $M_{l_1,l_2,m}(t)$ very similar to Eq. (20). Moreover, the dipole-dipole interaction contribution to the relaxation of the magnetization of a two spin (or magnetic particle) system can also be treated. Here, the reduced free energy $\beta V$ in the dc magnetic field may be written as[19]

$$\beta V = -\xi_\gamma (\cos\vartheta_1 + \cos\vartheta_2) - \sigma(\cos^2\vartheta_1 + \cos^2\vartheta_2)$$
$$- \frac{\mu^2}{kTr^3}[\cos(\vartheta_1 - \vartheta_2) - 3\cos(\vartheta_1 - \vartheta)\cos(\vartheta_2 - \vartheta)], \quad (36)$$

where $\vartheta$ is the bond angle and $r$ is the distance between two particles. Equation (36) is very similar to Eq. (13) thus the magnetization dynamics can also be studied using our method. In principle, a generalization of the approach developed to three, four, etc. spin systems is also straightforward, i.e., it is needed to modify the reduced free energy $\beta V$, Eq. (13) (by adding corresponding terms), and the stochastic moment functions in Eq. (16) (which will now include products of three, four, etc. spherical harmonics). However, the arrangement of resulted recurrence equation comprising a large number of indices into the tridiagonal form Eq. (24) becomes more and more complicated.

## IV. RESULTS AND DISCUSSION

The *linear response* relaxation time of $M_Z(t)$ evaluated from Eqs. (25) and (27), where $\Delta h = h_\mathrm{I} - h_\mathrm{II}$ is maintained at a small constant value 0.001, is shown in Fig. 1 as a function of the interaction (exchange coupling) parameter $\varsigma$ for *small* anisotropy $\sigma$ and $\alpha=1$ (IHD regime). In that Figure, Curve 1 represents the behavior of the relaxation time as a function of the exchange parameter for zero reduced bias field $h_\mathrm{II}$, Curves 2 and 3 represent the behavior of the relaxation time in linear response for finite values of the reduced bias field $h_\mathrm{II}$. Apparently the effect of increasing exchange coupling is to increase the relaxation time. The foregoing result pertains to relatively small anisotropy, which in the absence of exchange coupling is always treated by perturbation theory as described by Brown.[2,3] Figure 2 on the other hand essentially displays the effect of the exchange coupling for zero reduced bias field on the behavior of the relaxation time for all values of $\sigma$ (which may be considered as inverse temperature parameter as well). We remark that for $\sigma \geq 2$ in the absence of exchange coupling the relaxation time is exponentially large and is given by the inverse Kramers escape rate as described by Brown.[2,3] It is apparent from Fig. 2 that the effect of increasing exchange coupling in the zero bias field situation in linear response is again to increase the relaxation time, which is entirely in accord with curve 1 of Fig. 1. Although the present results are based on a two body interaction they at least allow one to understand qualitatively the role played by exchange interaction which is in general to increase the relaxation time over and above that in the absence of interaction. Fig. 3 displays the behavior of the relaxation time as a function of the exchange coupling for three reasonably large values of the anisotropy parameter $\sigma$ and a small bias field with $\Delta h$ still maintained very small so that we are again considering the linear response in the presence of a small bias field. As before the effect of increasing exchange coupling is to increase the relaxation time, that is, to raise the potential barrier height. In addition a comparison of the exact (numerical) integral relaxation time with the asymptotes obtained in Ref. 23 for the IHD situation typified by $\alpha = 1$ by means of Langer's theory, and summarized in Appendix B, is given in Figs. 2 and 3. First of all, these plots show overall good agreement between the asymptotic calculation and those rendered by the MCF method, excluding the low barrier region characterized by small $\sigma$ and exchange coupling values as exemplified by curve 1 in Fig.3. The deviation between the exact and asymptotic solutions for the relaxation time for small $\sigma$, or equivalently, low anisotropy energy barriers, is as expected because Langer's theory on which the asymptotic solution is based assumes high energy barriers and thus well-defined saddle points and metastable states. On examination of Fig. 3, one notices the striking result that the apparent discontinuity in the asymptotic expression for the relaxation time as a function of the exchange coupling (region between the dots and triangles) observed in Ref. 23 is not reproduced by the exact numerical MCF solution. Thus the discontinuity reported there is merely an artifact of the approximation used in the asymptotic calculation, i.e., the Taylor series expansion of the potential energy near the saddle point, as discussed in Ref. 23. The discrepancy between the exact and asymptotic solutions manifests itself in the vicinity of the critical exchange coupling. There the saddle points become rather flat so that Taylor series expansion of the potential energy near the saddle point is no longer valid. Therefore, the numerical solution indicates, despite the fact that a critical exchange coupling marking a fundamental change in the energyscape and therefore in the reversal mechanism (which changes from a two-step into a one-step process) exists, that no corresponding singularity in the relaxation time as a function of that coupling appears.

Thus in Figs. 2 and 3 the reasons for seeking the exact solution for the relaxation time become apparent. Namely the exact solution based on the MCF method yields the behavior of the relaxation time both in the region where the magnetic anisotropy energy is comparable to the thermal energy $kT$ and in the critical exchange coupling region where the saddle points flatten. Thus the exact solution allows one to accurately delineate the regions in which the asymptotic solution is no longer applicable. In other words, just as in the absence of exchange coupling, it is now possible to give a range of validity for the asymptotic formulae. Another advantage of the MCF solution is that it is valid for all $\alpha$ unlike the asymptotic solution based on Langer's method, which applies to intermediate-to-high damping only.

Figs 2 and 3 also indicate that for small $h_{II}$, the dependence of $\tau$ on the anisotropy parameter $\sigma$ has an activation character leading to *exponential growth* of the relaxation time $\tau$ as the height of the potential barrier $\sigma$ increases as predicted by transition state theory. However, as the constant bias field increases, so that, taking zero exchange coupling described by bistable potential as an example, the wells of the potential become markedly non equivalent and the integral relaxation time $\tau$ can *decrease* with *increasing* $\sigma$ (see Fig. 4). Thus $\tau$ may *differ exponentially* from the inverse Kramers rate. This effect was first reported in Ref. 30 and explained qualitatively in Ref. 31 in an analysis of the linear response of an assembly of noninteracting uniaxial particles in the low temperature limit. It is due to the depletion of the population of the upper or shallower potential well consequent on the escape of many particles from that well and their subsequent descent to the deeper well where it is very difficult to escape from due to the high energy barrier. In particular, the depletion effect is typified by the fact that for values of the parameters $h_I \cong h_{II}$ above a certain critical level $h_c \cong 0.17$, $\tau$ no longer has an activation character. Thus the integral relaxation time can no longer provide an accurate approximation to the reversal time of the magnetization. In other words in this situation contrary to the inverse Kramers rate or overbarrier relaxation time $\tau$ *decreases* as the height of the potential barrier *increases*. The depletion effect also occurs in the presence of exchange interaction and is indeed reinforced by that interaction as shown in Fig.4. Hence the integral relaxation time again *exponentially diverges* from the greatest relaxation time or inverse many body IHD Kramers escape rate yielded by Langer's method for values of the reduced field in excess of the critical value $h_c$.

The imaginary $\chi''(\omega)$ part of the complex susceptibility for, $\alpha =1$, $\sigma = 6$, $h_I =0.001$, $h_{II} =0$, and various values of the exchange parameter $\varsigma$ are shown in Fig. 5. Here two bands appear in the magnetic loss $\chi''(\omega)$ spectra. One relaxation band dominates the low-frequency part of the spectra and is due to the slow overbarrier relaxation of the spins. The second band is due to high frequency "intrawell" modes. Just as for noninteracting particles (Ref. 26, Chapters 7 and 9), the dynamic susceptibility $\hat{\chi}(\omega)$ given as an infinite series of Lorenzians may be approximated by a sum of two Lorentzians only

$$\hat{\chi}(\omega) \approx \frac{1-\Delta}{1+i\omega/\lambda_1} + \frac{\Delta}{1+i\omega\tau_W}. \qquad (37)$$

Here $\tau_W$ is the effective relaxation time characterizing the high-frequency "intrawell" modes and so has a weak temperature dependence, $\Delta$ is a parameter accounting their contribution to the spectrum (for small $h_{\text{II}}$, instead of the longest relaxation time $1/\lambda_1$ the integral relaxation time $\tau \approx 1/\lambda_1$ may be used). The parameters $\Delta$ and $\tau_W$ in Eq. (37) may be determined from $1/\lambda_1$, $\tau$, and $\tau^{ef}$ in such way as to guarantee the correct asymptotic behavior of $\chi(\omega)$ in the extreme cases of very low and very high frequencies (see for detail Ref. 26, Chap. 2). However, to fit the spectra in Fig. 5, we have used in Eq. (37) the only adjustable parameter $\Delta$ as here $\Delta \ll 1$ and $1/\lambda_1 \approx \tau$ so that an approximation $\tau_W \approx \Delta \tau^{ef}$ may be used (to satisfy approximately the asymptotic behavior of $\hat{\chi}(\omega)$ in the extreme cases of very low and very high frequencies). The times $\tau$ and $\tau^{ef}$ evaluated numerically from Eqs. (27) and (35), respectively. The low and very high frequency asymptotes, Eq. (8), are also shown for comparison. As apparent by inspection of curves in Fig. 5, the low frequency band shifts to lower frequencies as $\varsigma$ increases (in accordance with the results shown in Figs. 2 and 3). We remark that for small values of $\alpha$ the third (ferromagnetic resonance) peak appears in the high frequency part of the spectrum $\chi''(\omega)$ due to the precessional motion of the spins.

Table 1. Numerical values of parameters in Eq. (37) used in Fig. 5.

| $\varsigma$ | $\tau/\tau_N$, Eq. (27) | $\tau^{ef}/\tau_N$, Eq. (35) | $\Delta$, the best fit |
|---|---|---|---|
| 0.01 | 61.1 | 10.52 | 0.01 |
| 1.0 | 143.8 | 18.65 | 0.005 |
| 5.0 | 3348 | 29.14 | 0.003 |

The presentation of the dynamic susceptibility $\hat{\chi}(\omega)$ by Eq. (37) implies that the long time behavior of $M_Z(t)$ may be approximated by two exponentials corresponding to overbarrier and "intrawell" relaxation processes (on neglecting the contribution of precessional modes)

$$M_Z(t)/M_Z(0) \approx (1-\Delta)e^{-t\lambda_1} + \Delta e^{-t/\tau_W}. \qquad (38)$$

Equation (38) allows one to readily estimate the dynamics of $M_Z(t)$ in the time domain (for $t > \tau_N$ the contribution of the second term may be ignored).

To summarize, in this paper we have presented a general method for the calculation of the response functions and relaxation times of the magnetization for linear and nonlinear transient responses of two interacting spins using the matrix continued fraction technique including nonlinear effects *for all* values of anisotropy and interaction parameters. Simple approximate equations for the dynamic susceptibility and the time behavior of the magnetization are also given. It appears that including the ferromagnetic exchange interaction always tends to increase the effective barrier height, that is, to increase the greatest relaxation time. Moreover, it appears that the depletion effect where the integral relaxation time may diverge exponentially from the inverse Kramers rate consequent on the application of a strong bias field is enhanced by the exchange interaction. The calculations we have presented also allow one to pinpoint the critical exchange coupling regions where calculations based on Langer's method become invalid. In particular, we have proved that the discontinuity which appears in the relaxation time in a certain critical region of the exchange coupling predicted by the asymptotic solution is simply an artifact of that solution. We emphasize that the asymptotic solution based on Langer's method is valid for the IHD region only where the energy loss per cycle of a spin with the capability to cross a barrier is much greater than *kT*. No such restriction of course applies to the exact solution which in turn suggests that the asymptotic solution should be generalized to include all values of $\alpha$ and then tested against the exact solution. Such a generalization has been made for non interacting spins in Ref. 32, where a single asymptotic formula for the greatest relaxation time including the low damping, Kramers turnover and IHD regions has been given by generalizing the Mel'nikov-Meshkov solution of the Kramers turnover problem for particles to spins. This calculation could be extended to include the effect of exchange coupling.

**ACKNOWLEDGMENTS**

The support of this work by INTAS (project 01-2341) is gratefully acknowledged. S.V.T. thanks the University of Versailles for the award of a Visiting Professorship.

**APPENDIX A. MATRICES $\mathbf{Q}_n, \mathbf{Q}_n^+, \mathbf{Q}_n^-$ AND INITIAL VALUE VECTORS $\mathbf{C}_n(0)$**

The matrices $\mathbf{Q}_n, \mathbf{Q}_n^+, \mathbf{Q}_n^-$ have the form

$$\mathbf{Q}_n^- = \begin{pmatrix} \mathbf{V}_{2n-1} & \mathbf{R}_{2n-1} \\ \mathbf{0} & \mathbf{V}_{2n} \end{pmatrix}, \quad \mathbf{Q}_n = \begin{pmatrix} \mathbf{P}_{2n-1} & \mathbf{S}_{2n-1} \\ \mathbf{R}_{2n} & \mathbf{P}_{2n} \end{pmatrix}, \quad \mathbf{Q}_n^+ = \begin{pmatrix} \mathbf{U}_{2n-1} & \mathbf{0} \\ \mathbf{S}_{2n} & \mathbf{U}_{2n} \end{pmatrix} \quad (A1)$$

where

$$\mathbf{P}_m = \begin{pmatrix} \mathbf{p}_{m,0} & \mathbf{p}_{0,m}^* & \ddots & & \mathbf{0} \\ \mathbf{p}_{m-1,1}^* & \mathbf{p}_{m-1,1} & \ddots & & \ddots \\ \ddots & \ddots & \ddots & & \mathbf{p}_{m-1,1}^* \\ \mathbf{0} & & \ddots & \mathbf{p}_{0,m}^* & \mathbf{p}_{0,m} \end{pmatrix},$$

$$\mathbf{R}_m = \begin{pmatrix} \mathbf{r}_{0,m}^* & \mathbf{0} & \ddots & & \mathbf{0} \\ \mathbf{r}_{m-1,1} & \mathbf{r}_{1,m-1}^* & \ddots & & \ddots \\ \mathbf{0} & \mathbf{r}_{m-2,2} & \ddots & & \mathbf{0} \\ \ddots & \ddots & \ddots & & \mathbf{r}_{m-1,1}^* \\ \mathbf{0} & & \ddots & \mathbf{0} & \mathbf{r}_{0,m} \end{pmatrix},$$

$$\mathbf{S}_m = \begin{pmatrix} \mathbf{s}_{m,0} & \mathbf{s}_{0,m}^* & \mathbf{0} & \ddots & \mathbf{0} \\ \mathbf{0} & \mathbf{s}_{m-1,1} & \mathbf{s}_{1,m-1}^* & \ddots & \ddots \\ \ddots & \ddots & \ddots & \ddots & \mathbf{0} \\ \mathbf{0} & \ddots & \mathbf{0} & \mathbf{s}_{0,m} & \mathbf{s}_{m,0}^* \end{pmatrix},$$

$$\mathbf{V}_m = \begin{pmatrix} \mathbf{v}^*_{0,m} & \mathbf{0} & \ddots & & \mathbf{0} \\ \mathbf{v}^*_{m-1,1} & \mathbf{v}^*_{1,m-1} & \ddots & & \ddots \\ \mathbf{v}^*_{m-2,2} & \mathbf{v}_{m-2,2} & \ddots & & \mathbf{0} \\ \mathbf{0} & \mathbf{v}^*_{m-3,3} & \ddots & & \mathbf{v}^*_{m-2,2} \\ \ddots & \ddots & \ddots & & \mathbf{v}_{1,m-1} \\ \mathbf{0} & \ddots & \mathbf{0} & & \mathbf{v}^*_{0,m} \end{pmatrix},$$

$$\mathbf{U}_m = \begin{pmatrix} \mathbf{u}^*_{m,0} & \mathbf{u}_{m,0} & \mathbf{u}^*_{0,m} & \mathbf{0} & \ddots & \mathbf{0} \\ \mathbf{0} & \mathbf{u}^*_{m-1,1} & \mathbf{u}_{m-1,1} & \mathbf{u}^*_{1,m-1} & \ddots & \ddots \\ \ddots & \ddots & \ddots & \ddots & \ddots & \mathbf{0} \\ \mathbf{0} & \ddots & \mathbf{0} & \mathbf{u}^*_{0,m} & \mathbf{u}_{0,m} & \mathbf{u}^*_{m,0} \end{pmatrix}.$$

The matrices $\mathbf{p}_{n,m}$, $\mathbf{s}_{n,m}$, $\mathbf{u}^*_{n,m}$, $\mathbf{v}^*_{n,m}$ have the form

$$\mathbf{x}_{n,m} = \begin{pmatrix} \ddots & \ddots & \ddots & \ddots & 0 \\ \ddots & x_{n,m,-1} & 0 & \ddots & \ddots \\ \ddots & 0 & x_{n,m,0} & 0 & \ddots \\ \ddots & \ddots & 0 & x_{n,m,1} & \ddots \\ 0 & \ddots & \ddots & \ddots & \ddots \end{pmatrix}_{(2r+1)\times(2r_x+1)}.$$

The matrices $\mathbf{p}^*_{n,m}$, $\mathbf{r}_{n,m}$, $\mathbf{u}_{n,m}$, $\mathbf{v}_{n,m}$ are given by

$$\mathbf{x}_{n,m} = \begin{pmatrix} \ddots & \ddots & \ddots & \ddots & 0 \\ \ddots & x_{n,m,-1} & x^+_{n,m,-1} & 0 & \ddots \\ \ddots & x^-_{n,m,0} & x_{n,m,0} & x^+_{n,m,0} & \ddots \\ \ddots & 0 & x^-_{n,m,1} & x_{n,m,1} & \ddots \\ 0 & \ddots & \ddots & \ddots & \ddots \end{pmatrix}_{(2r+1)\times(2r_x+1)}.$$

Here $\mathbf{x}$ designates one of the submatrices $\mathbf{p}_{n,m}$, $\mathbf{p}^*_{n,m}$, $\mathbf{r}_{n,m}$, $\mathbf{s}_{n,m}$, $\mathbf{u}_{n,m}$, $\mathbf{u}^*_{n,m}$, $\mathbf{v}_{n,m}$, $\mathbf{v}^*_{n,m}$. All the submatrices has the same number of rows, namely, $2r+1$, where $r = \min[n,m]$. The number of columns also can be found as $2r_x + 1$, but now each submatrix has its own number $r_x$, namely

$$r_p = \min[n,m], \quad r_{p^*} = \min[n+1,m-1], \quad r_s = \min[n+1,m], \quad r_r = \min[n,m-1],$$

$$r_v = \min[n-1,m-1], \quad r_u = \min[n+1,m+1], \quad r_{v^*} = \min[n,m-2], \quad r_{u^*} = \min[n+2,m].$$

The corresponding matrix elements are

$$p_{l_1,l_2,m} = d_{l_1,l_2,m}^{l_1,l_2,m} = -\sum_{l=l_1,l_2}\left(\frac{1}{2}l(l+1) - \sigma\frac{l(l+1)-3m^2}{(2l-1)(2l+3)}\right),$$

$$p^*_{l_1,l_2,m} = d_{l_1+1,l_2-1,m}^{l_1,l_2,m} = \frac{1}{2}\varsigma(l_2-l_1+1)\sqrt{\frac{((l_1+1)^2-m^2)(l_2^2-m^2)}{(2l_1+1)(2l_1+3)(2l_2-1)(2l_2+1)}},$$

$$p^{*\pm}_{l_1,l_2,m} = d_{l_1+1,l_2-1,m\pm1}^{l_1,l_2,m} = \frac{1}{4}\varsigma(l_2-l_1+1)\sqrt{\frac{(l_1\pm m+1)(l_1\pm m+2)(l_2\mp m-1)(l_2\mp m)}{(2l_1+1)(2l_1+3)(2l_2-1)(2l_2+1)}},$$

$$s_{l_1,l_2,m} = d_{l_1+1,l_2,m}^{l_1,l_2,m} = -\left(\frac{\xi_{\text{II}}}{2}l_1 + \frac{i(2\sigma-\varsigma)}{2\alpha}m\right)\sqrt{\frac{(l_1+1)^2-m^2}{4(l_1+1)^2-1}},$$

$$s^*_{l_1,l_2,m} = d_{l_1+1,l_2,m\pm1}^{l_1,l_2,m} = \pm\frac{i\varsigma}{4\alpha}\sqrt{\frac{(l_1\pm m+1)(l_1\pm m+2)(l_2\pm m+1)(l_2\mp m)}{(2l_1+1)(2l_1+3)}},$$

$$r_{l_1,l_2,m} = d_{l_1,l_2-1,m}^{l_1,l_2,m} = \left(\frac{\xi_{\text{II}}}{2}(l_2+1) + \frac{i(2\sigma-\varsigma)}{2\alpha}m\right)\sqrt{\frac{l_2^2-m^2}{4l_2^2-1}},$$

$$r^*_{l_1,l_2,m} = d_{l_1,l_2-1,m\pm1}^{l_1,l_2,m} = \pm\frac{i\varsigma}{4\alpha}\sqrt{\frac{(l_1\pm m+1)(l_1\mp m)(l_2\mp m-1)(l_2\mp m)}{(2l_2-1)(2l_2+1)}},$$

$$u_{l_1,l_2,m} = d_{l_1+1,l_2+1,m}^{l_1,l_2,m} = -\frac{1}{2}\varsigma(l_1+l_2)\sqrt{\frac{((l_1+1)^2-m^2)((l_2+1)^2-m^2)}{(2l_1+1)(2l_1+3)(2l_2+1)(2l_2+3)}},$$

$$u^{\pm}_{l_1,l_2,m} = d_{l_1+1,l_2+1,m\pm1}^{l_1,l_2,m} = \frac{1}{4}\varsigma(l_1+l_2)\sqrt{\frac{(l_1\pm m+1)(l_1\pm m+2)(l_2\pm m+1)(l_2\pm m+2)}{(2l_1+1)(2l_1+3)(2l_2+1)(2l_2+3)}},$$

$$u^*_{l_1,l_2,m} = d_{l_1+2,l_2,m}^{l_1,l_2,m} = -\sigma\frac{l_1}{2l_1+3}\sqrt{\frac{((l_1+1)^2-m^2)((l_1+2)^2-m^2)}{(2l_1+1)(2l_1+5)}},$$

$$v_{l_1,l_2,m} = d_{l_1-1,l_2-1,m}^{l_1,l_2,m} = \frac{1}{2}\varsigma(l_1+l_2+2)\sqrt{\frac{(l_1^2-m^2)(l_2^2-m^2)}{(2l_1-1)(2l_1+1)(2l_2-1)(2l_2+1)}},$$

$$v^{\pm}_{l_1,l_2,m} = d_{l_1-1,l_2-1,m\pm1}^{l_1,l_2,m} = -\frac{1}{4}\varsigma(l_1+l_2+2)\sqrt{\frac{(l_1\mp m-1)(l_1\mp m)(l_2\mp m-1)(l_2\mp m)}{(2l_1-1)(2l_1+1)(2l_2-1)(2l_2+1)}},$$

$$v^*_{l_1,l_2,m} = d_{l_1,l_2-2,m}^{l_1,l_2,m} = \sigma\frac{l_2+1}{2l_2-1}\sqrt{\frac{(l_2^2-m^2)((l_2-1)^2-m^2)}{(2l_2+1)(2l_2-3)}},$$

and

$$d^{l_i,l_j,m}_{l_i+x,l_j+y,m\pm1} = \left(d^{l_j,l_i,m}_{l_j+y,l_i+x,m\pm1}\right)^*.$$

The initial value vectors $\mathbf{C}_n(0)$ in Eq. (25) are calculated in the following manner. We introduce the vector

$$\mathbf{F}^\gamma_n = \begin{pmatrix} \mathbf{f}^\gamma_{2n-1,0} \\ \mathbf{f}^\gamma_{2n-2,1} \\ \vdots \\ \mathbf{f}^\gamma_{0,2n-1} \\ \mathbf{f}^\gamma_{2n,0} \\ \mathbf{f}^\gamma_{2n-1,1} \\ \vdots \\ \mathbf{f}^\gamma_{0,2n} \end{pmatrix}_{4n^2+2n+1} \qquad \mathbf{f}^\gamma_{n,m} = \begin{pmatrix} M^\gamma_{n,m,-r} \\ M^\gamma_{n,m,-r+1} \\ \vdots \\ M^\gamma_{n,m,r} \end{pmatrix}, \quad r = \min[n,m],$$

where the index $\gamma =$ I, II corresponds to the fields $\mathbf{H}^I_Z$ and $\mathbf{H}^{II}_Z$. Next, we transform Eq. (22) to the matrix recursion formula

$$\mathbf{Q}^-_n \mathbf{F}^\gamma_{n-1} + \mathbf{Q}_n \mathbf{F}^\gamma_n + \mathbf{Q}^+_n \mathbf{F}^\gamma_{n+1} = \mathbf{0}.$$

The solution of this equation has the form

$$\mathbf{F}^\gamma_n = \mathbf{\Delta}^\gamma_n(0) \mathbf{Q}^-_n \mathbf{F}^\gamma_{n-1} = \frac{1}{4\pi} \mathbf{\Delta}^\gamma_n(0) \mathbf{Q}^-_n \mathbf{\Delta}^\gamma_{n-1}(0) \mathbf{Q}^-_n \ldots \mathbf{\Delta}^\gamma_1(0) \mathbf{Q}^-_1.$$

Here, we have noted that $\mathbf{F}^\gamma_0 = 1/(4\pi)$. Thus we can write the initial value vector as

$$\mathbf{C}_n(0) = \mathbf{F}^I_n - \mathbf{F}^{II}_n.$$

**APPENDIX B. ESCAPE RATE OF A SYSTEM OF TWO INTERACTING SPINS**

The longest relaxation time for the TSP in the IHD regimes ($\alpha \geq 1$) can be estimated in terms of the relaxation rate $\Gamma$ as [23]

$$\tau/\tau_N \quad \left[\Gamma(h_\gamma) + \Gamma(-h_\gamma)\right]^{-1}, \tag{B1}$$

$$\Gamma = \begin{cases} 2\Gamma^+\Gamma^-/(\Gamma^+ + \Gamma^-) & j < 1 - h^2_\gamma \\ 2\Gamma_{2s} & j > 1 - h^2_\gamma \end{cases}, \tag{B2}$$

where

$$\Gamma_{2s} = \frac{\sigma^{3/2}(1-h_\gamma^2)(1-h_\gamma)[j+1-h_\gamma]}{2\sqrt{\pi j[j-(1-h_\gamma^2)]}} e^{-2\sigma(1-h_\gamma)^2},$$

$$\Gamma^\pm = \frac{\sigma^{3/2}|\kappa^\pm|D^\pm P^\pm}{\sqrt{\pi j|2R_1^\pm R_2^\pm + jW^\pm(R_1^\pm + R_2^\pm)|}} e^{-\beta\Delta V^\pm},$$

$$P^\pm = [(1-\cos^2\vartheta_1^\pm)(1-\cos^2\vartheta_2^\pm)]^{1/4},$$

$$R_{1,2}^\pm = \cos\vartheta_{1,2}^\pm(2\cos\vartheta_{1,2}^\pm + h_\gamma) - 1,$$

$$W^\pm = P^\pm - j/2 \pm h_\gamma\sqrt{1+(j/2)^2/(1-j)},$$

$$\beta\Delta V^\pm = -\sigma\left\{ j\sqrt{\left[1\pm h_\gamma\sqrt{1+(j/2)^2/(1-j)} - j/2\right]^2 - \left(h_\gamma \pm \sqrt{1-j}\right)^2}\right.$$
$$\left. + j\left[1\pm h_\gamma\sqrt{1+(j/2)^2/(1-j)} - j/2\right] - \left[h_\gamma^2 + j - 1 \pm 2h_\gamma\sqrt{1+(j/2)^2/(1-j)}\right] - 2 + 2h_\gamma \pm (2h_\gamma - j)\right\},$$

$$D^+ = (1-h_\gamma)(j+1-h_\gamma), \quad D^- = 1-h_\gamma^2 - j,$$

$j = \varsigma/\sigma$. Here $\cos\vartheta_{1,2}^\pm$ corresponds to the position of the saddle points (two saddle points for each spin) which can be found from the equations $\beta\frac{\partial V}{\partial\vartheta_{1,2}} = 0$ so that

$$\cos\vartheta_1^\pm = \frac{1}{2}\left(-h_\gamma \mp \sqrt{1-j} + \sqrt{\left(h_\gamma \pm \sqrt{1-j}\right)^2 + 2j \mp 4h_\gamma\sqrt{1+(j/2)^2/(1-j)}}\right),$$

$$\cos\vartheta_2^\pm = \frac{1}{2}\left(-h_\gamma \mp \sqrt{1-j} - \sqrt{\left(h_\gamma \pm \sqrt{1-j}\right)^2 + 2j \mp 4h_\gamma\sqrt{1+(j/2)^2/(1-j)}}\right).$$

The attempt frequencies $\kappa^\pm$ are computed numerically.[23]

The relaxation time for the two-spin system contains two unconnected branches corresponding to the two regimes, $j < j_c = 1-h_\gamma^2$ and $j > j_c$;[23] these branches are plotted in Figs. 2 and 3 by filled circles and triangles, respectively. Apparently a "critical" value of the exchange coupling, $j_c = 1-h_\gamma^2$ exists separating two regimes with distinct reversal mechanisms. More precisely, for strong coupling ($j > j_c$) the two spins reverse their direction coherently through the only available saddle point. On the other hand, for weak coupling two saddle points exist so that reversal of the two spins is a two-step process. Moreover, near $j_c$ it was shown that the relaxation rate diverges due to flattening of the saddle points implying that Langer's general calculation of the IHD relaxation rate which is formulated by means of a quadratic expansion of the potential energy about the saddle point, is not applicable in this coupling range, thus leaving the two ranges unconnected as far as asymptotic calculations are concerned.

# REFERENCES

1. L. Néel, Ann. Géophys. **5**, 99 (1949).

2. W. F. Brown, Jr., Phys. Rev. **130**, 1677 (1963).

3. W. F. Brown, Jr., IEEE Trans. Mag. **15**, 1196 (1979).

4. W. Wernsdorfer, Adv. Chem. Phys. **118**, 99 (2001).

5. J. P. Chen, C. M. Sorensen, K. J. Klabunde, and G. C. Hadjipanayis, Phys. Rev. B **51**, 11527 (1995).

6. M. Respaud, J. M. Broto, H. Rakoto, A. R. Fert, L. Thomas, et al., Phys. Rev. B **57**, 2925 (1998).

7. R. H. Kodama and A. E. Berkowitz, Phys. Rev. B **59**, 6321 (1999).

8. J. T. Richardson, D. I. Yiagas, B. Turk, and J. Forster, J. Appl. Phys. **70**, 6977 (1991).

9. A.J. Budó, Chem. Phys. **17**, 686 (1949).

10. W.T. Coffey, M.W. Evans, and P. Grigolini, *Molecular diffusion and spectra* (Wiley, New York, 1984).

11. A. Liberatos, E. P. Wohlfarth, and R. W. Chantrell, IEEE Trans. Magn. **MAG-21**, 1277 (1985).

12. A. Liberatos and R. W. Chantrell, J. Appl. Phys. **73**, 6501 (1993).

13. D. Hinzke and U. Nowak, Phys. Rev. B **61**, 6734 (2000);

14. D. Hinzke and U. Nowak, J. Magn. Magn. Mater. **221**, 365 (2000).

15. S. I. Denisov and K. N. Trohidou, Phys. Rev. B **64**, 184433 (2001).

16. J. L. Dormann, L. Bessais, and D. Fiorani, J. Phys. C **21**, 2015 (1988).

17. S. Mørup and E. Tronc, Phys. Rev. Lett. **72**, 3278 (1994).

18. P. E. Jonsson and J. L. Garcia-Palacios, Europhys. Lett. **55**, 418 (2001).

19. D. Rodé, H. N. Bertram, and D. R. Fradkin, IEEE Trans. Magn. **MAG-23**, 2224 (1987).

20. W. Chen, S. Zhang, and H. N. Bertram, J. Appl. Phys. **71**, 5579 (1992).

[21] A. Yoshimori and J. Korringa, Phys. Rev. **128**, 1054 (1962); J. Korringa and A. Yoshimori, Phys. Rev. **128**, 1060 (1962).

[22] I. Solomon, Phys. Rev. **99**, 559 (1955).

[23] H. Kachkachi, Europhys. Lett. **62** (5), 650 (2003).

[24] J. S. Langer, Ann. Phys. (N.Y.) **54**, 258 (1969).

[25] H. B. Braun, J. Appl. Phys. **76**, 6310 (1994).

[26] W. T. Coffey, Yu. P. Kalmykov, and J. T. Waldron, *The Langevin Equation*, 2nd Ed. (World Scientific, Singapore, 2004).

[27] W. T. Coffey, D. A. Garanin, and D. J. McCarthy, Adv. Chem. Phys. **117**, 528 (2001).

[28] H. Risken, *The Fokker-Planck Equation*, 2nd Edition (Springer, Berlin, 1989).

[29] D. A. Varshalovich, A. N. Moskalev, and V. K. Khersonskii, *Quantum Theory of Angular Momentum* (World Scientific, Singapore, 1998).

[30] W. T. Coffey, D. S. F. Crothers, Yu. P. Kalmykov, E. S. Massawe, and J. T. Waldron, Phys. Rev. E **49**, 1869 (1994).

[31] D. A. Garanin, Phys. Rev. E **54**, 3250 (1996).

[32] Yu. P. Kalmykov, W. T. Coffey, and S. V. Titov, Fizika Tverdogo Tela **47**, 260 (2005) [Phys. Solid State, **47**, 272 (2005)].

**FIGURE CAPTIONS**

**Figure 1.** Relaxation time $\tau/\tau_N$ vs. $\varsigma$ for $\sigma=1$, $\alpha=1$, and various values of $h_{\text{II}}$ in the linear response condition $h_{\text{I}} - h_{\text{II}} = 0.001$.

**Figure 2.** Relaxation time $\tau/\tau_N$ vs. $\sigma$ for $\alpha=1$, $h_{\text{I}} = 0.001$, $h_{\text{II}} = 0$ and various values of $\varsigma$. Exact matrix continued fraction solution for the relaxation time $\tau/\tau_N$ [solid lines: Eq.(27))] is compared with the inverse reaction rate rendered by Langer's theory of the decay of metastable states [filled circles and triangles: Eqs. (B1) and (B2)].

**Figure 3.** Relaxation time $\tau/\tau_N$ vs. $\varsigma$ for $\alpha=0.5$, $h_{\text{I}} = 0.101$, $h_{\text{II}} = 0.1$, and various values of $\sigma$. Exact solution [solid lines: Eq.(27)] is compared with the inverse reaction rate rendered by Langer's theory of the decay of metastable states [filled circles and triangles: Eq. (B1) and (B2)].

**Figure 4.** Relaxation time $\tau/\tau_N$ vs. $\sigma$ for $\alpha=1$, $h_{\text{I}} = 3.001$, $h_{\text{II}} = 3.0$, and various values of the exchange parameter $\varsigma$.

**Figure 5.** The imaginary part of the complex susceptibility $\chi''(\omega)$ vs. $\omega\tau_N$ for $\alpha=1$, $\sigma=6$, $h_{\text{I}} = 0.001$, $h_{\text{II}} = 0$, and various values of the exchange parameter $\varsigma = 0.01$, 1.0, and 5. Solid lines: the matrix continued fraction solution; filled circles: the approximate Eq. (37); dotted and dashed lines: the low and high frequency asymptotes Eq. (8), respectively.

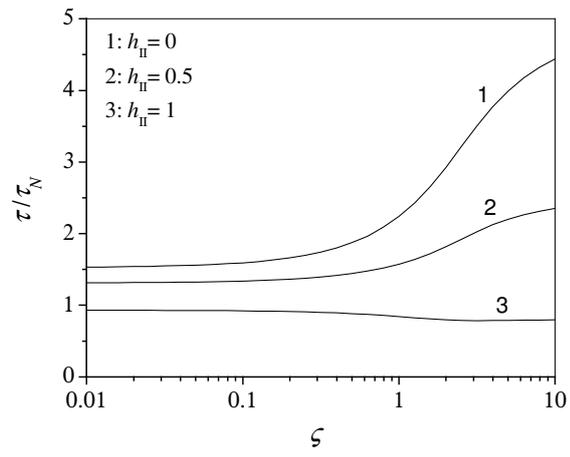

Fig.1

(Titov et al.)

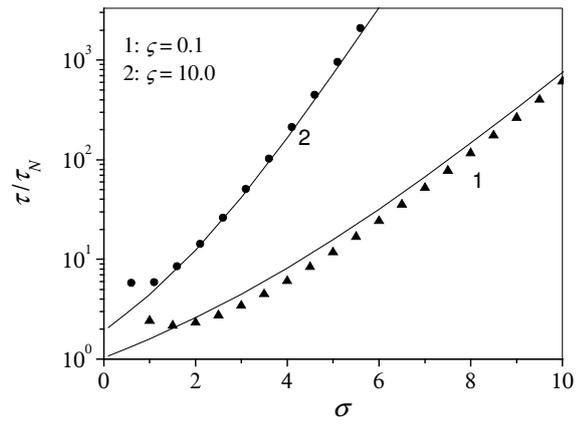

Fig. 2

(Titov et al.)

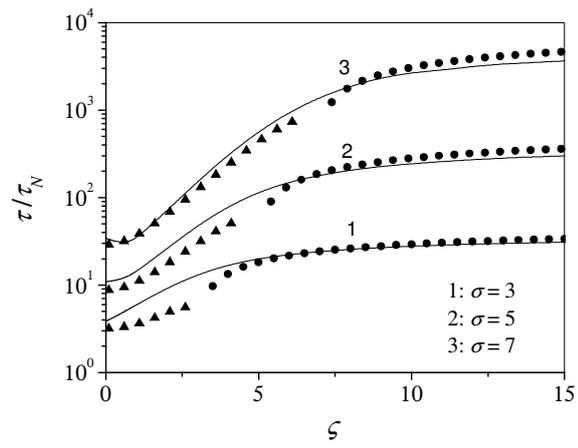

Fig. 3

(Titov et al.)

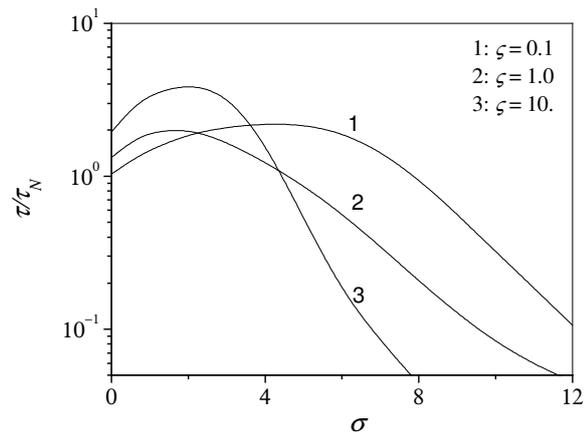

Fig.4

(Titov et al.)

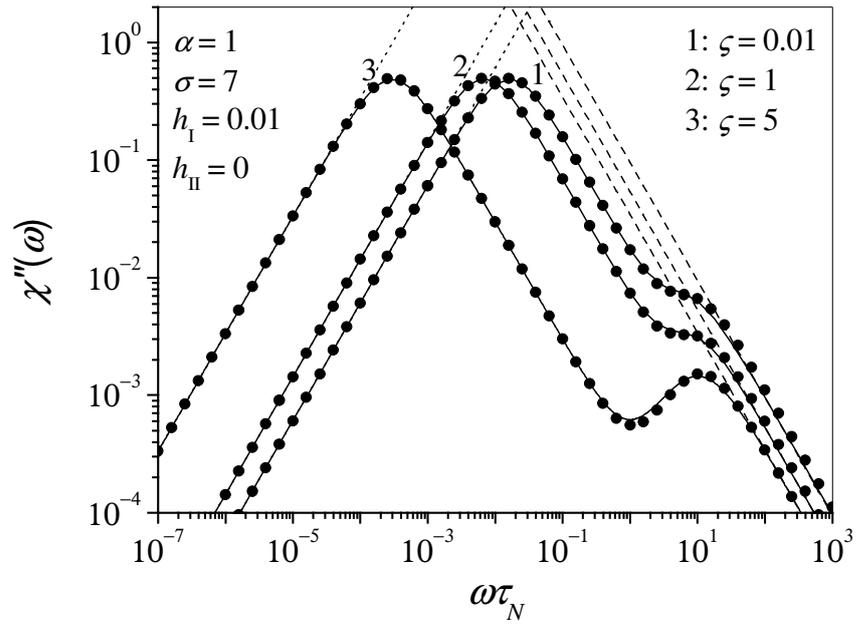

Fig. 5

(Titov et al.)